\documentstyle[aps,prl,twocolumn,epsf]{revtex}
\newcommand{\beq}{\begin{equation}}
\newcommand{\eeq}{\end{equation}}
\newcommand{\bea}{\begin{eqnarray}}
\newcommand{\eea}{\end{eqnarray}}

\begin{document}

\title{ Shell correction energy for bubble nuclei }

\author{ Yongle YU$^{1}$, Aurel BULGAC $^{1}$ and Piotr MAGIERSKI $^{1,2,3}$}

\address{ $^1$Department of Physics,  University of Washington,
Seattle, WA 98195--1560, USA  }
\address{$^2$ The Royal Institute of Technology, Physics Department Frescati,
Frescativ\"agen 24, S--10405, Stockholm, SWEDEN }
\address{$^3$ Institute of Physics, Warsaw University of Technology,
ul. Koszykowa 75, PL--00662, Warsaw, POLAND }

\date{\today }

\maketitle

\begin{abstract}

The positioning of a bubble inside a many fermion system does not
affect the volume, surface or curvature terms in the liquid drop
expansion of the total energy. Besides possible Coulomb effects, the
only other contribution to the ground state energy of such a system
arises from shell effects. We show that the potential energy surface
is a rather shallow function of the displacement of the bubble from
the center and in most cases the preferential position of a bubble is
off center. Systems with bubbles are expected to have bands of
extremely low lying collective states, corresponding to various bubble
displacements.

\end{abstract}




\vspace{0.5cm}

\narrowtext

There are a number of situations when the formation of voids is
favored. When a system of particles has a net charge, the Coulomb
energy can be significantly lowered if a void is created
\cite{wheeler,pomorski} and despite an increase in surface energy the
total energy decreases.  One can thus naturally expect that the
appearance of bubbles will be favored in relatively heavy nuclei. This
situation has been considered many times over the last 50 years in
nuclear physics and lately similar ideas have been put forward for
highly charged alkali metal clusters \cite{dietrich}.

The formation of gas bubbles is another suggested mechanism which
could lead to void(s) formation \cite{moretto}. The filling of a
bubble with gas prevents it from collapsing.  Various heterogeneous
atomic clusters \cite{saito} and halo nuclei \cite{austin} can be
thought of as some kind of bubbles as well.  In these cases, the
fermions reside in a rather unusual mean--field, with a very deep well
near the center of the system and a very shallow and extended one at
its periphery.  Since the amplitude of the wave function in the
semiclassical limit is proportional to the inverse square root of the
local momentum, the single--particle (s.p.) wave functions for the weakly
bound states will have a small amplitude over the deep well. If the
two wells have greatly different depths, the deep well will act almost
like a hard wall (in most situations).

Several aspects of the physics of bubbles in Fermi systems have not
been considered so far in the literature.  It is tacitly assumed that
a bubble position has to be determined according to symmetry
considerations. For a Bose system one can easily show that a bubble
has to be off--center \cite{chin}.  In the case of a Fermi system the
most favorable arrangement is not obvious \cite{bubbles}. The total
energy of a many fermion system has the general form
\begin{equation}
E(N)=e_vN +e_sN^{2/3}+e_cN^{1/3} + E_{sc}(N),
\label{eq:liq}
\end{equation}
where the first three terms represent the smooth liquid drop part of
the total energy and $E_{sc}$ is the pure quantum shell correction
contribution, the amplitude of which grows in magnitude approximately
as $\propto N^{1/6}$, see Ref. \cite{strutin}.  We shall consider in
this work only one type of fermions with no electric charge. In a
nuclear system the Coulomb energy depends rather strongly on the
actual position of the bubble, but in a very simple way.  In an alkali
metal cluster, as the excess charge is always localized on the
surface, the Coulomb energy is essentially independent of the bubble
position.  The character of the shell corrections is in general
strongly correlated with the existence of regular and/or chaotic
motion \cite{balian,strut}. If a spherical bubble appears in a
spherical system and if the bubble is positioned at the center, then
for certain ``magic'' fermion numbers the shell correction energy
$E_{sc}(N)$, and hence the total energy $E(N)$, has a very deep
minimum. However, if the number of particles is not ``magic'', in
order to become more stable the system will in general tend to deform.
Real deformations lead to an increased surface area and liquid drop
energy.  On the other hand, merely shifting a bubble off--center
deforms neither the bubble nor the external surface and therefore, the
liquid drop part of the total energy of the system remains unchanged.

Moving the bubble off--center can often lead to a greater stability of
the system due to shell correction energy effects.  In recent years it
was shown that in a 2--dimensional annular billiard, which is the
2--dimensional analog of spherical bubble nuclei, the motion becomes
more chaotic as the bubble is moved further from the center
\cite{bohigas}.  One might thus expect that the importance of the
shell corrections diminishes when the bubble is off--center. We shall
show that this is not the case however.

One can anticipate that the relative role of various periodic orbits
(diameter, triangle, square etc.) is modified in unusual ways in
systems with bubbles.  In 3D--systems the triangle and
square orbits determine the main shell structure and produce the
beautiful supershell phenomenon \cite{balian,ben}. A small bubble
near the center will affect only diameter orbits. After being
displaced sufficiently far from the center, the bubble will first
touch and destroy some triangle orbits. In a 3D--system
only a relatively small fraction of these orbits will be destroyed.
Thus one might expect that the existence of supershells will not be
critically affected, but that the supershell minimum will be less
pronounced.  A larger bubble will simultaneously affect triangular and
square orbits, and thus can have a dramatic impact on both shell and
supershell structure.
\begin{figure}[thb]
\begin{center}
\epsfxsize=7.25cm
\centerline{\epsffile{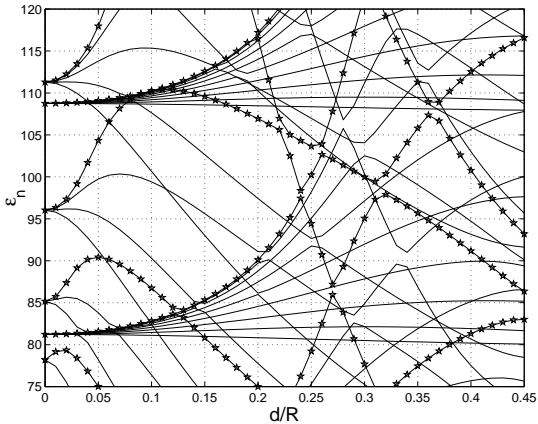}}
\end{center}
\caption{A portion of the full unfolded s.p. spectrum (with unit average
level density) for the case of a bubble of radius $a= R/2$
($R=R_0N^{1/3}$) as a function of the bubble displacement $d/R$. 
Energy levels with $m=0$ (single--degenerate) are marked with pentagrams.}
\label{fig1}
\end{figure}

The change of the total energy of a many fermion system can be
computed quite accurately using the shell corrections method, once the
s.p. spectrum is known as a function of the shape of the
system \cite{strutin,strut}. The results presented in this Letter have
been obtained using the 3D--version of the conformal mapping method
described in \cite{bubbles} as applied to an infinite square well
potential with Dirichlet boundary conditions. The magic numbers are
hardly affected by the presence or absence of a small diffuseness
\cite{lew}. The absence of a spin--orbit interaction leads to
quantitative, but to no qualitative differences. 

In Fig. 1 we show the unfolded s.p. spectrum for the case of a bubble
of half the radius of the system, $a=R/2$, as a function of the
displacement $d/R$ of the bubble from the center. The size of the
system is determined as usual from $R^3-a^3=r_0^3N$.  The unfolded
s.p. spectrum is determined using the Weyl formula \cite{weyl} for the
average cumulative number of states.
\begin{equation}
\varepsilon _n = N_{W}(e_n),
\label{eq:sp}
\end{equation}
where $e_n$ are the actual s.p. energies of the Schr\"odinger
equation, $N_{W}(e)$ is the Weyl formula for the total number of
states with energy smaller than $e$ in a 3D--cavity and $\varepsilon
_n$ are the unfolded eigenvalues, which by construction leads to a
spectrum with an unit average level density.  As the bubble is moved
off center, the classical problem becomes more chaotic \cite{bohigas}
and one can expect that the s.p. spectrum would approach that of a
random Hamiltonian \cite{gian} and that the nearest--neighbor
splitting distribution would be given by the Wigner
surmise\cite{mehta}. A random Hamiltonian would imply that ``magic''
particle numbers are as a rule absent.  There is a large number of
avoided level crossings in Fig. 1 and one can clearly see a
significant number of relatively large gaps in the spectrum. Note that
levels with different symmetries (different angular momentum
projection on the symmetry axis $m$) can cross.  Even for extreme
displacements large gaps in the s.p. spectrum occur significantly more
frequently than in the case of a random (which is closer to an uniform)
spectrum. A simple estimate, using the Wigner surmise, shows that gaps
of the order of 3 units or larger should be absent in the portion of
the spectrum shown in Fig. 1. The probability to encounter a nearest
neighbor energy spacing $s$ greater than $s_0$ is given by
$P(s>s_0)=\exp ( -\pi s_0^2/4 )$.  For $s_0=3,4,5$ one thus obtains $
8.5\times 10^{-4}$, $3.5\times 10^{-6}$ and $3\times 10^{-9}$
respectively.  Several very large gaps for $d/R\approx 0.45$ are
unambiguously present. Higher in the spectrum even larger gaps could
be found.  These features are definitely not characteristic of a
random Hamiltonian.  If the particle number is such that the Fermi
level is at a relatively large gap, then the system at the
corresponding ``deformation'' is very stable. A simple inspection of
Fig. 1 suggests that for various particle numbers the energetically
most favorable configuration can either have the bubble on-- or
off--center. This situation is very similar to the celebrated
Jahn--Teller effect in molecules.  Consequently, a ``magic'' particle
number could correspond to a ``deformed'' system. In this respect this
situation is a bit surprising, but not unique. It is well known that
many nuclei prefer to be deformed, and there are particularly stable
deformed ``magic'' nuclei or clusters \cite{strut,ben,lew,bohr}.

There is a striking formal analogy between the energy shell correction
formula and the recipe for extracting the renormalized vacuum Casimir
energy in quantum field theory \cite{cas} or the critical Casimir
energy in a binary liquid mixture near the critical demixing point
\cite{fish}. Note that even though Casimir energy is typically a
smooth function of distance, it cannot be ascribed to the ``smooth
liquid drop'' energy. Similarly, no part of the $E_{sc}$ energy of a
bubble near the surface can be ascribed to the ``smooth liquid
drop'' energy. In Fig. 2 we show the contour plot of the $E_{sc}$
energy for a system with $a=R/2$ as a function of the bubble
displacement $d/R$ versus $N^{1/3}$. The overall regularity of
``mountain ridges' and ``canyons'' seem to be due the interference
effects arising from two periodic orbits along the diameter passing
through the centers of the two spheres. Various mountain tops and
valleys form an alternating network almost orthogonal to the
``mountain ridges'' and ``canyons''. For some $N$'s the bubble
``prefers'' to be in the center, while for other values that is the
highest energy configuration.

As a function of the particle number $N$ and at fixed $d/R$, the
oscillation amplitude of the shell correction energy is maximal for
on--center configurations.  For a given particle number $N$ the energy
is an oscillating function of the displacement $d$ and many
configurations at different $d$ values have similar energies. However,
in all cases, moving the bubble all the way to the edge of the system
leads to the lowest values of $E_{sc}(N)$. This drop in the shell
correction energy as a function of $d$ is preceded by the highest
``mountain range''. A practitioner of the 
\begin{figure}[thb]
\begin{center}
\epsfxsize=7.25cm
\centerline{\epsffile{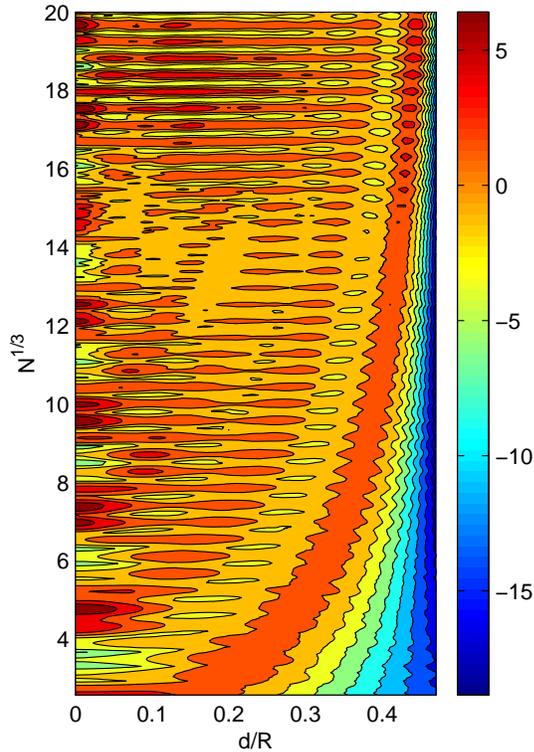}}
\end{center}
\caption{Contour plot of the shell correction energy for the case of a
bubble of radius $a = R/2$ for up to $N=8,000$ spinless
fermions. Energy is measured in units of $\hbar^2/2mr_0^2$.}
\label{fig2}
\end{figure}
\noindent Strutinsky method might be
tempted to ascribe these features to the smooth part of the total
energy. One should remember however that the Strutinsky recipe
requires a smearing energy $\gamma$, which is supposed to be chosen
larger than the typical energy separation between two consecutive
energy shells. In a semiclassical language, such a difference is
determined by the shortest periodic orbit in the system. In the
present case the length of the shortest orbit $2(R-d-a)\rightarrow 0$,
when the bubble approaches the edge of the system. This would require
an ever longer smearing interval $\gamma$ in order to perform the
Strutinsky procedure.  In the absence of analytical results for this
system a comparison with a simpler situation is extremely
illuminating. When the inner and outer surfaces are very close one can
ignore in the first approximation their curvatures and consider instead
the case of matter between two infinite parallel planes. It can be
shown explicitly that the shell correction energy is inversely
proportional to the separation between the two surfaces \cite{bm}, a
behavior which is similar to that seen in Fig. 2. For a small bubble
one can easily agree that it is more cost effective to make a hole
closer to the edge, where the s.p.w.fs. are smaller.  Once again, we
note here the analogy with the Casimir energy \cite{cas,fish}.
Moreover, at least qualitatively, this shortest orbit and the one
diametrically opposed to it suffice to explain the pattern of
``valleys'' and ``ridges'' in Figs. 2 and 3.  It is not entirely clear
to us whether this final drop in the total energy could occur in a
self--sustaining system. When the bubble is close to the outer
surface, matter density in the region of the closest approach
decreases, which in turn leads to a decrease of the self--consistent
potential. In this case the square well potential model used by us
becomes then inadequate. Physical systems where such configurations
can nevertheless be realized are briefly mentioned at the end.  In the
case of a bubble with a smaller radius $a=R/5$ the number of level
crossings is significantly smaller than in Fig. 1. As a result, the
shell correction energy contour plot has less structure, see Fig. 3,
and thus a system with a smaller bubble is also significantly softer.
\begin{figure}[thb]
\begin{center}
\epsfxsize=7.25cm
\centerline{\epsffile{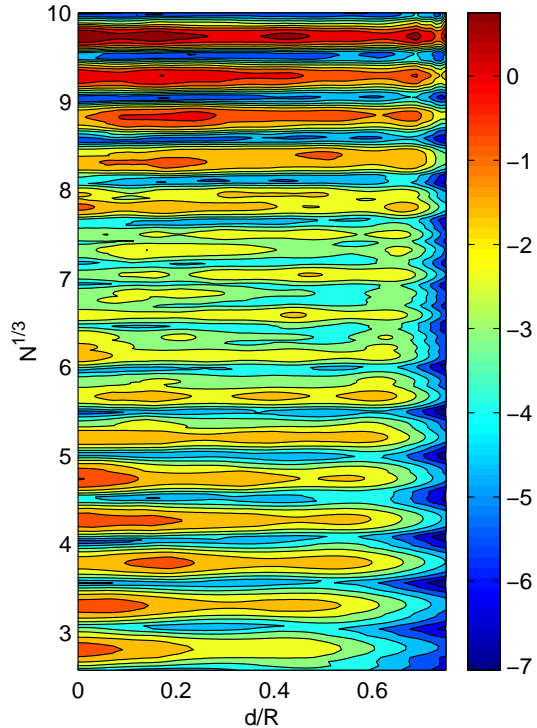}}
\end{center}
\caption{The same as in Fig. 2 but for $a = R/5$ and for up to
$N=1,000$ spinless fermions.}
\label{fig3}
\end{figure}

Pairing correlations can lead to a further softening of the potential
energy surface of a system with one or more bubbles. We have seen that
the energy of a system with a single bubble is an oscillating function
of the bubble displacement. When the energy of the system as a
function of this displacement has a minimum, the Fermi level is in a
relatively large gap, where the s.p. level density is very low. When
the energy has a maximum, just the opposite takes place.  Pairing
correlations will be significant when the Fermi level occurs in a
region of high s.p. level density and it is thus natural to expect
that the total energy is lowered by paring correlations at ``mountain
tops'', and be less affected at ``deep valleys". All this ultimately
leads to a further leveling of the potential energy surface.  With
increasing temperature the shell correction energy decreases in
magnitude, but the most probable position of a bubble is still
off--center. The reason in this case is however of a different nature,
the ``positional'' entropy of such a system favors configurations with
the bubble off-center, as a simple calculation shows, namely
$S_{pos}({\bf d}) = 2\ln d +{\rm const}$, where ${\bf d}$ is the
position vector of the center of the bubble with respect to the center
of the sphere.  Moreover, making more bubbles could lead to a further
decrease of the free energy, even though the total energy might
increase. 

A system with one or several bubbles should be a very soft system.
The energy to move a bubble is parametrically much smaller than any
other collective mode. All other familiar nuclear collective modes for
example involve at least some degree of surface deformation.  For this
reasons, once a system with bubbles is formed, it could serve as an
extremely sensitive ``measuring device'', because a weak external
field can then easily perturb the positioning of the bubble(s) and
produce a system with a completely different geometry. There are quite
a number of systems where one can expect that the formation of bubbles
is possible \cite{bubbles}.  Known nuclei are certainly too small and
it is difficult at this time to envision a way to create nuclei as big
as those predicted in Refs. \cite{pomorski}. On the other hand voids,
not always spherical though, can be easily conceived to exist in
neutron stars \cite{chris}. Metallic clusters with bubbles, one or
more fullerenes in a liquid metal or a metallic ball placed inside a
superconducting microwave resonator \cite{caio} in order to study the
ball energetics and maybe even dynamics, are all very promising
candidates.

Financial support for this research was provided by DOE, the Swedish
Institute and the G\"oran Gustafsson Foundation. We thank O. Bohigas
for his lively interest and for reading the draft and making a few
suggestions and S.A. Chin and H.A. Forbert for discussions.


\vspace{-0.75cm}

\begin{thebibliography}{99}

\vspace{-0.25cm}

\bibitem{wheeler} H.A. Wilson, Phys. Rev. {\bf 69} 538 (1946);
J.A. Wheeler, unpublished notes; P.J. Siemens and H.A. Bethe,
Phys. Rev. Lett. {\bf 18}, 704 (1967); C.Y. Wong, Ann. Phys.  {\bf
77}, 279 (1973); W.J. Swiatecki, Physica Scripta {\bf 28}, 349 (1983);
W.D. Myers and W.J. Swiatecki, Nucl. Phys. {\bf A 601}, 141 (1996).

\bibitem{pomorski} K. Pomorski and K. Dietrich, Nucl. Phys. {\bf A
627}, 175(1997); Phys. Rev. Lett. {\bf 80}, 37 (1998); J. Decharg\'e
{\it et al.}, Phys. Lett. {\bf B 451}, 275 (1999).

\bibitem{dietrich} K. Pomorski and K. Dietrich, Eur. Journ. Phys. {\bf
D 4}, 353 (1998).

\bibitem{moretto} L.G. Moretto {\it et al.}, Phys. Rev. Lett. {\bf
78}, 824 (1997).

\bibitem{saito} S. Saito and F. Yabe, in {\it
Chemistry and Physics of Fullerenes and Related Materials}, vol. {\bf
6}, eds. K.M. Kadish and R.S. Ruoff, Pennington, 1998, pp 8--20; T.P. Martin
{\it et al.}, J. Chem. Phys. {\bf 99}, 4210 (1993); U. Zimmermann {\it
et al.}, Phys. Rev. Lett.  {\bf 72}, 3542 (1994).

\bibitem{austin} S.M. Austin and G.F. Bertsch, Scientific American,
{\bf 272}, 62 (1995).

\bibitem{chin} S.A. Chin and H.A. Forbert, Los Alamos e--preprint
archive, cond--mat/9810269.

\bibitem{bubbles} A. Bulgac {\it et al.}, in Proc. of the
Int. Workshop on {\it Collective excitations in Fermi and Bose
systems''}, eds. C.A. Bertulani, L.F. Canto and M.S. Hussein,
pp. 44--61, World Scientific, Singapore (1999); Los Alamos e--preprint
archive, nucl-th/9811028.

\bibitem{strutin} V.M. Strutinsky and A.~G. Magner,
Sov. J. Part. Nucl. Phys. {\bf 7}, 138 (1976).

\bibitem{balian} R. Balian and C. Bloch, Ann. Phys. {\bf 67}, 229
(1972); M. Brack and R.K. Bhaduri, {\it Semiclassical Physics},
Addison--Wesley, Reading, MA (1997).

\bibitem{strut} V.M. Strutinsky, Sov. J. Nucl. Phys. {\bf 3}, 449
(1966); Nucl.  Phys. {\bf A 95}, 420 (1967); {\it ibid }{\bf A 122}, 1
(1968); M. Brack {\it et al.}, Rev. Mod. Phys. {\bf 44}, 320 (1972).

\bibitem{bohigas} O. Bohigas {\it et al.}, Phys. Rep.  {\bf 223}, 43
(1993); O. Bohigas {\it et al.}, Nucl. Phys. A {\bf 560}, 197 (1993);
S. Tomsovic and D. Ullmo, Phys. Rev. E {\bf 50}, 145 (1994);
S.D. Frischat and E. Doron, Phys. Rev. E {\bf 57}, 1421 (1998).

\bibitem{ben} H. Nishioka {\it et al.}, Phys. Rev. B {\bf 42}, 9377
(1990); J. Pedersen {\it et al.}, Nature {\bf 353}, 733 (1991);
M. Brack, Rev. Mod. Phys. {\bf 65}, 677 (1993) and references therein.

\bibitem{lew} A. Bulgac and C. Lewenkopf, Phys. Rev. Lett. {\bf 71},
4130 (1993).

\bibitem{weyl} R.T. Waechter, Proc. Camb.  Phil. Soc. {\bf 72}, 439
(1972); H.P. Baltes and E.R. Hilf, {\it Spectra of Finite Systems},
Wissenschaftsverlag, Mannheim, Wien, Z\"urich: Bibliographisches
Institut, (1976).

\bibitem{gian} O. Bohigas {\it et al.}, Phys. Rev. Lett. {\bf 52}, 1
(1984).

\bibitem{mehta} M.L. Mehta, {\it Random Matrices}, Academic Press
Inc., Boston, 1991.

\bibitem{bohr} A. Bohr and B. Mottelson, {\it Nuclear Structure},
vol. II, Benjamin, New York, (1974).

\bibitem{cas} H.B.G. Casimir, Proc. K. Ned. Akad. Wet. {\bf 51}, 793
(1948); V.M. Mostepanenko and N.N. Trunov, Sov. Phys. Usp. {\bf 31},
965 (1988) and references therein.

\bibitem{fish} M.E. Fisher and P.G. de Gennes, C.R. Acad. Sci. Ser. B
{\bf 287}, 207 (1978); A. Hanke {\it et al.}, Phys. Rev. Lett. {\bf
{81}}, 1885 (1998) and references therein.

\bibitem{bm} A. Bulgac and P. Magierski, unpublished.

\bibitem{chris} G. Baym {\it et al.}, Nucl. Phys. {\bf A175}, 225
(1971); C.P. Lorentz {\it et al.}, Phys. Rev. Lett. {\bf 70}, 379
(1993); C.J. Pethick and D.G. Ravenhall,
Ann. Rev. Nucl. Part. Sci. {\bf 45}, 429 (1995) references therein;
H. Heiselberg {\it et al.}, Phys. Rev. Lett. {\bf 70}, 1355 (1993).

\bibitem{caio} H.D. Graf {\it et al.}, Phys. Rev. Lett. {\bf 69}, 1296
(1992).

\end{thebibliography}
\end{document}